\documentclass[
    ,final            
  ]
  {aipproc}

\layoutstyle{6x9}

\begin{document}

\title{Predictions of Warped Extra Dimensions for Flavor Phenomenology}


\author{Stefania Gori}{
  address={Physik Department, Technische Universit\"at M\"unchen,
D-85748 Garching, Germany\\
Max-Planck-Institut f{\"u}r Physik (Werner-Heisenberg-Institut), 
D-80805 M{\"u}nchen, Germany}
}

\begin{abstract}
The aim of these proceedings is to present the main predictions of the Randall-Sundrum model with custodial protection for particle-antiparticle mixing and rare decays of $K$ and $B_{s,d}$ mesons, putting particular attention on the testability of the resulting NP effects at future experiments. Before giving numerical results, we discuss theoretical expectations, residing in the flavor structure of the model. The high energy scale $M_{KK}$ is chosen in such a way that direct searches of new particles at the LHC are possible, 
still being consistent with electroweak precision observables.
\end{abstract}

\maketitle


\section{Introduction}
Randall-Sundrum (RS) models~\cite{Randall:1999ee}, in which all Standard Model (SM) fields are allowed to propagate in the bulk~\cite{Gherghetta:2000qt,Chang:1999nh,Grossman:1999ra}, represent a very exciting alternative to more
traditional extensions of the SM, like supersymmetry. In addition to a geometrical solution of the gauge hierarchy problem, among the achievements of this New Physics (NP) scenario, we mention the natural generation of hierarchies in fermion masses and weak mixing angles~\cite{Gherghetta:2000qt,Grossman:1999ra} and the suppression of Flavor Changing Neutral Current (FCNC) interactions~\cite{Huber:2003tu,Agashe:2004cp}.

In this class of models, once the value of the warping factor $e^{kL}$ is set in order to address the gauge hierarchy problem, the Kaluza Klein (KK) scale $M_{KK}$ is the only free parameter coming from geometry. In the simplest RS model with only the SM gauge group in the bulk, $M_{KK}$ has a quite strict bound ($\sim 10$ TeV), coming from the requirement of fitting the electroweak (EW) parameters. However a necessary, though not always sufficient, condition for direct signals of RS models at the LHC is the existence of KK modes with $\mathcal{O}(1\rm{ TeV})$ masses. To reduce the NP effects in the EW parameters as well as the NP contributions to the coupling $Zb_L\bar b_L$, the gauge group in the bulk is enlarged to $G=SU(3)\times SU(2)_L\times SU(2)_R\times U(1)\times P_{LR}$ (RS model with custodial protection) and the fermions are put in symmetric representations of $G$~\cite{Agashe:2003zs}\footnote{More specifically the left-handed down quarks are eigenstates of $P_{LR}$, in such a way that the coupling $Zb_L\bar b_L$ is SM-like.}. This allows for a scale $M_{KK}\sim (2-3)$TeV~\cite{Cacciapaglia:2006gp}. In this framework (see also~\cite{Albrecht:2009xr} for a detailed description) the flavor structure is quite rich, since the parameter space includes $18$ real parameters and $9$ phases in addition to the $(9+1)$ free parameters present in the SM~\cite{Blanke:2008zb}.

\section{The Flavor Structure}
The model goes beyond Minimal Flavor Violation (MFV). Basically there are two main origins of non-MFV effects:

\begin{itemize}
\item The explanation of the hierarchies of SM fermion masses and mixings leads to non-universalities in the KK gauge boson-SM fermion interaction, implying FCNCs at tree level mediated by the heavy gauge bosons. The same holds also for SM gauge bosons which, after electroweak Symmetry Breaking (EWSB), mix with the KK gauge bosons. Hence, also effects of non-unitarity of the CKM matrix appear~\cite{Buras:2009ka}.
\item The mixing, through EWSB, of the new heavy fermions and the SM ones~\cite{Buras:2009ka}\footnote{Effects of non-unitarity of the CKM matrix are resulting also from the mixing KK-SM fermions, even if these effects are generally smaller than those generated through the mixing between gauge bosons~\cite{Buras:2009ka}.}.
\end{itemize}

FCNCs at tree level usually constitute a serious problem of models beyond the Standard Model (BSM). However in the RS model with custodial protection the suppression of FCNC transitions to an acceptable level, even in presence of relatively light KK gauge bosons, can be guaranteed by two mechanisms: 
the {\it custodial protection} of all the flavor off-diagonal $Zd_L^i\bar d_L^j$ couplings~\cite{Blanke:2008zb}
and the {\it RS-GIM mechanism}~\cite{Agashe:2004cp}.

In the next section, we will show that these two protection mechanisms can not eliminate severe constraints coming from the CP violating (CPV) observable $\epsilon_k$.

 %
 \begin{center}
\begin{figure}[t]\label{fig:finetuning}
 \includegraphics[height=.15\textheight]{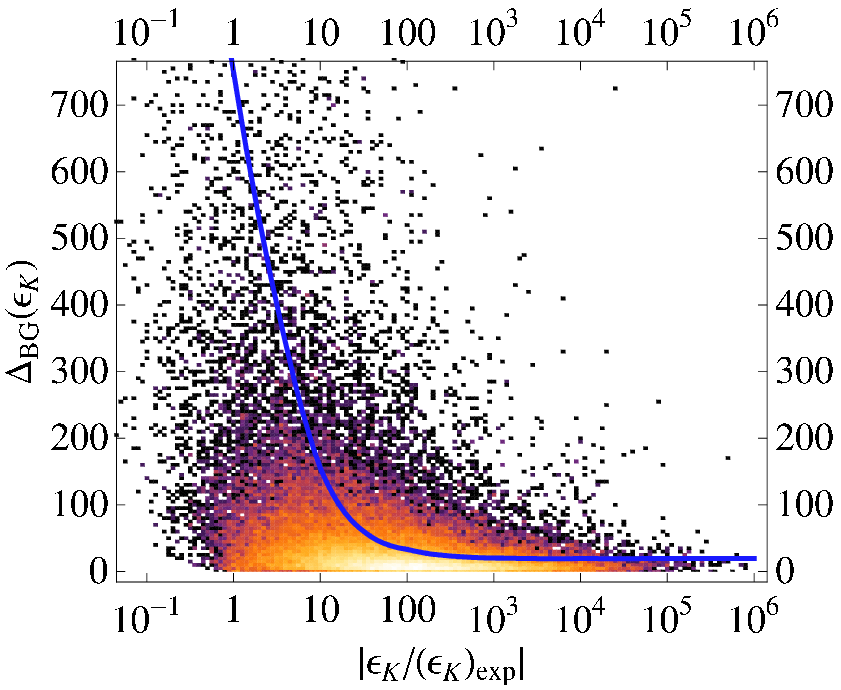}\hspace{0.3cm}
  \includegraphics[height=.15\textheight]{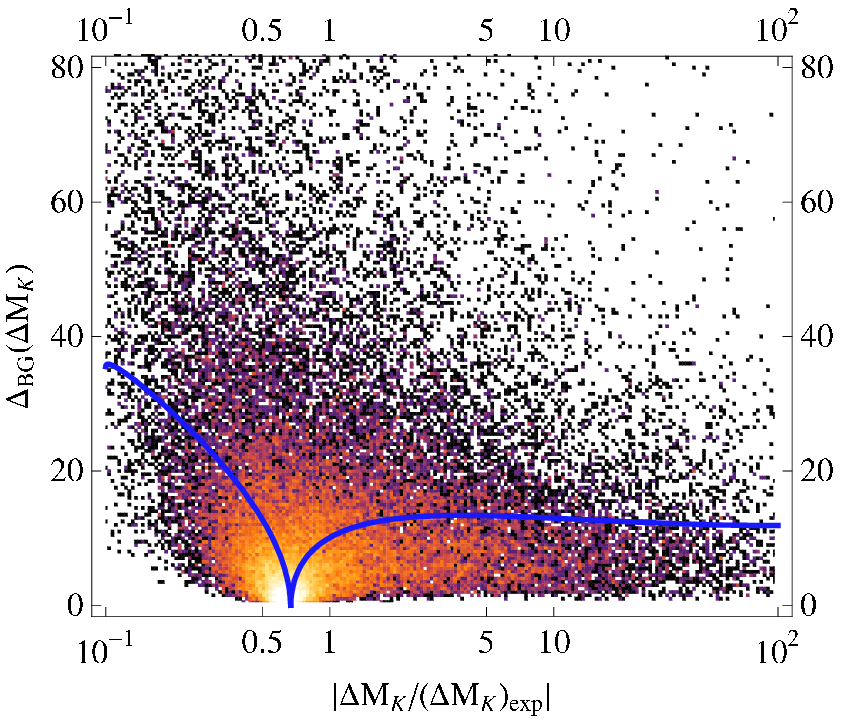}\hspace{0.3cm}
    \includegraphics[height=.15\textheight]{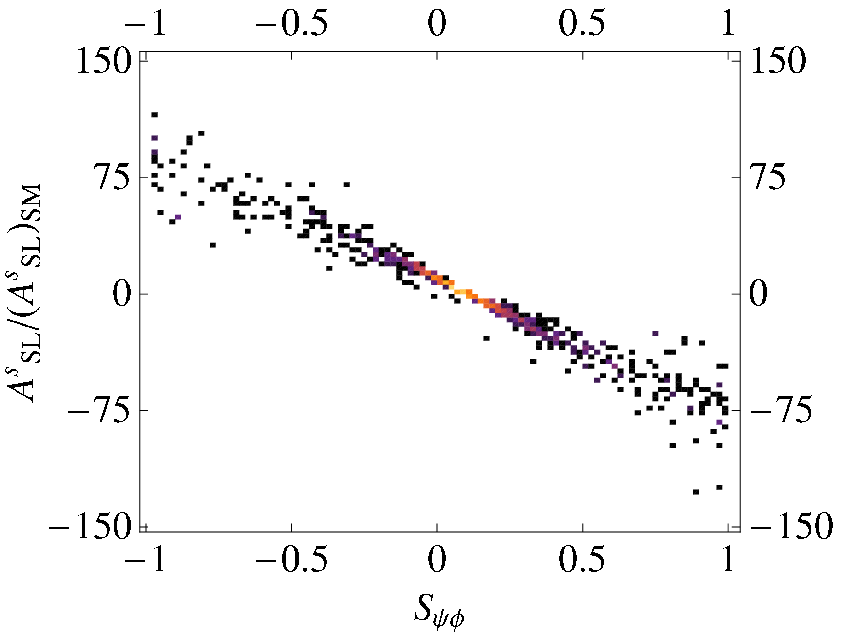}\hspace{0.3cm}
  \caption{Left: The fine-tuning $\Delta_{BG}(\epsilon_k)$ plotted against $\epsilon_K$ normalized to its experimental value. The blue line represents the average of fine-tuning required. Middle: The same for $\Delta M_K$. Right: Correlation between $S_{\psi\phi}$ and the semileptonic asymmetry $A_{SL}^s$ normalized to its SM value.}
\end{figure}
 \end{center}
\vspace{-1.4cm}
\section{Meson Mixing} 
Due to the presence of right-handed currents, $K$ and $B_{s,d}$ meson mixings receive a contribution not only from the operator already present in the SM,
but also from new operators, like the scalar left-right $Q_2^{LR}=\left(\bar q_1 P_L q_2\right)\left(\bar q_1 P_R q_2\right)$, generated by the exchange of KK-gluons at tree level.
 In the case of the $K$ system, this operator is strongly chirally and QCD enhanced. Assuming anarchic 5D Yukawas and light $M_{KK}$, these effects are too large to be controlled by the RS-GIM mechanism, resulting in a general too large NP contribution in $\epsilon_k$, as also shown in~\cite{Csaki:2008zd} where a lower bound on $M_{KK}$ of $\sim 20$ TeV was found. To further investigate the issue, we perform a fine tuning analysis~\cite{Blanke:2008zb} of $\epsilon_k$ (left panel of fig.~\ref{fig:finetuning}), finding that, because of the rich parameter space, there are points which are able to fit the experimental $\epsilon_k$. Among them $30\%$ require a fine tuning smaller than 20.
On the other hand, we do not expect a too large NP contribution for the CP conserving (CPC) observable $\Delta M_K$ (middle panel of fig.~\ref{fig:finetuning}). This is due to the fact that in the SM, differently from $\epsilon_K$, $\Delta M_K$ depends on the real part of the mixing amplitude $(M_{12}^K)_{SM}$ and not on its very small imaginary part. 

The observable $\epsilon_k$ turned out to be the only restrictive constraint, coming from meson mixings. In fact, in the $B_{s,d}$ systems the QCD and chiral enhancements of the operator $Q_2^{LR}$ are much smaller than in the $K$ system\footnote{Hence, the contributions of the EW gauge bosons $Z_H$, $Z^\prime$ and of the KK gluons are comparable~\cite{Blanke:2008zb}.}, resulting in a not too large NP contribution to the CPC and CPV observables.
Proceeding with the subset of phenomenologically valid points of the parameter space, we analyse 
the possible NP effects in observables, where a large room for NP is still open. In the right panel of fig.~\ref{fig:finetuning} we show that the model can easily account for the recent hints of NP in the CP asymmetry $S_{\psi\phi}$~\cite{Barberio:2008fa}.

 \begin{center}
\begin{figure}[t]\label{fig:rare}
 \includegraphics[height=.17\textheight]{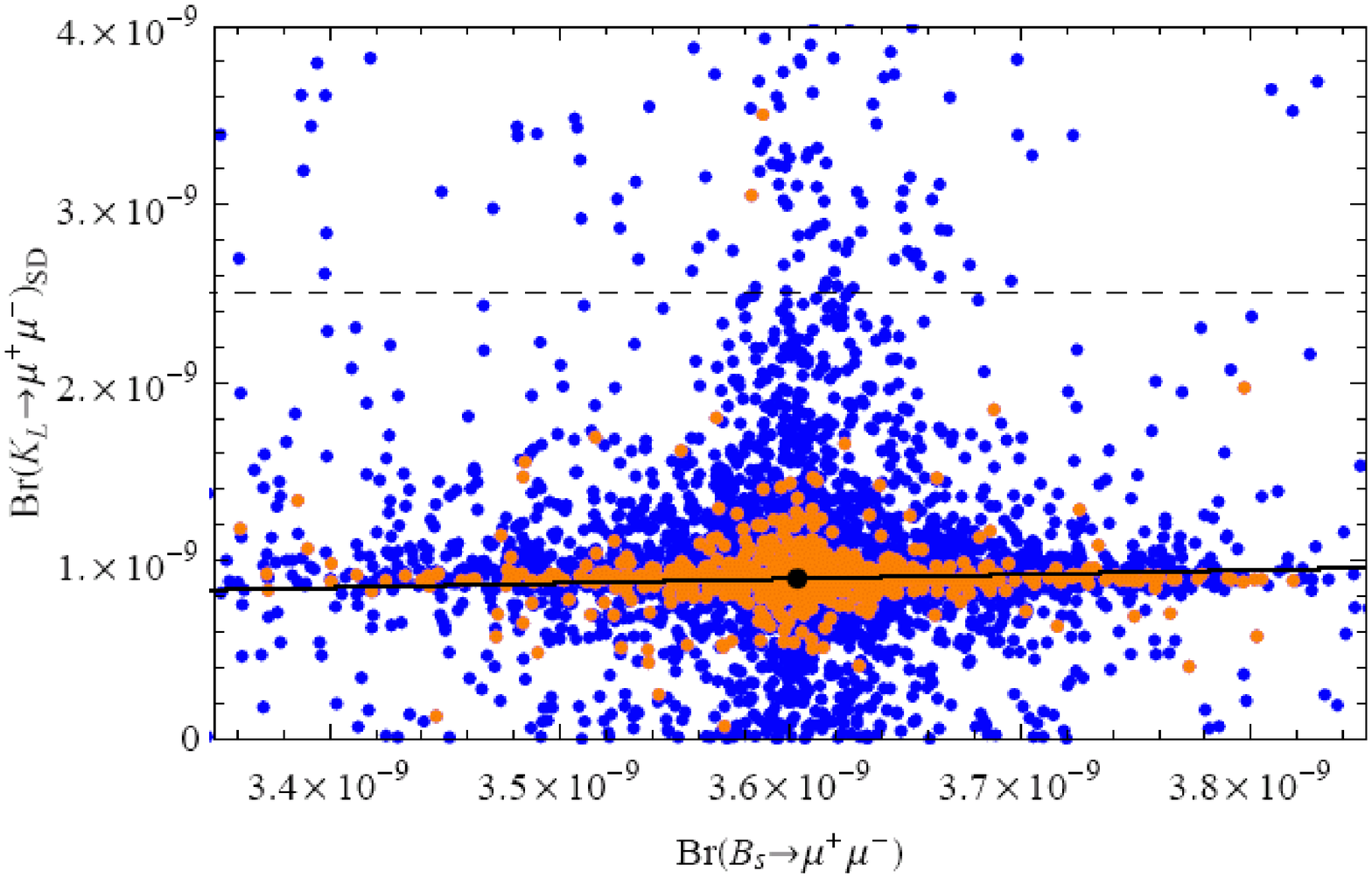}\hspace{0.3cm}
 \includegraphics[height=.17\textheight]{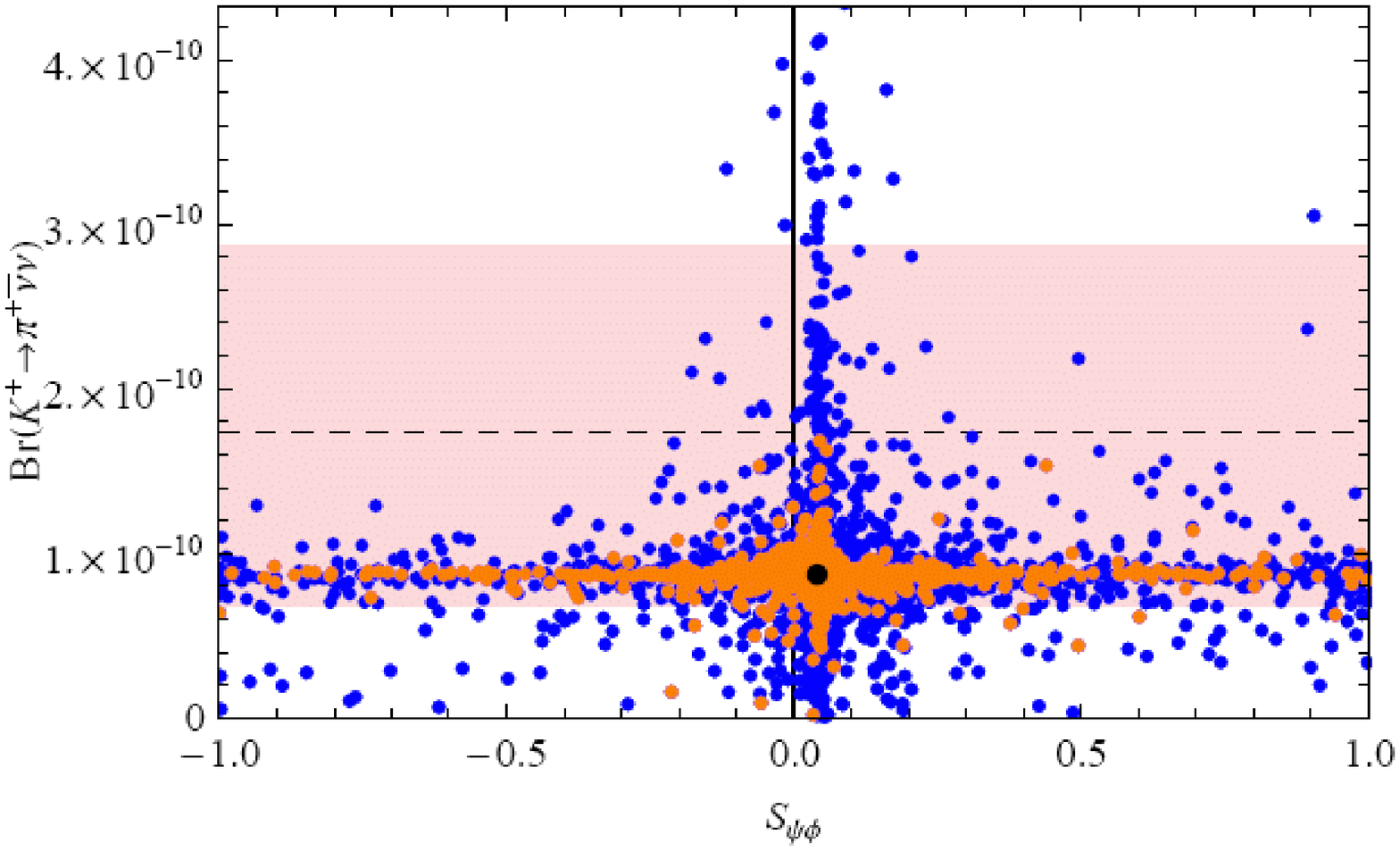}
  \caption{Left: Correlation between the short distance contribution to $K_L\rightarrow \mu^+\mu^-$ and $B_s\rightarrow \mu^+\mu^-$. The solid line represents the prediction of models with MFV. Right: Correlation between $K^+\rightarrow \pi^+\nu\bar\nu$ and $S_{\psi\phi}$. In orange the points which show a fine tuning $\Delta_{BG}(\epsilon_K)<20$. The black point is the SM prediction.}
\end{figure}
\end{center}

\vspace{-1.4cm}
\section{Rare Decays of $\mathbf K$ and $\mathbf{B_{s,d}}$ Mesons}

The $Z$ boson coupled with right-handed down quarks gives the main contribution~\cite{Blanke:2008yr} to the rare decays of $K$ and $B_{s,d}$ mesons
.
In contrast with the SM, where the $K$ and $B_{s,d}$ systems are governed by a flavor-universal loop function $X(x_t)$, the NP contributions are proportional to the CKM factors $1/\lambda_t^{(q)}$ ($q=K,d,s$), implying generally larger NP effects in $K$ decays than in $B_{s,d}$ decays, because of the hierarchy $\lambda_t^{(K)}\ll \lambda_t^{(d)}, \lambda_t^{(s)}$.

In order to reduce the parameter dependence, in fig.~\ref{fig:rare} we investigate some possible correlations between different observables (for a complete analysis see~\cite{Blanke:2008yr}). 
 
 \begin{itemize}
 \item $K_L\rightarrow \mu^+\mu^-$ vs $B_s\rightarrow \mu^+\mu^-$ (left panel): Due to the general much larger NP contributions in the rare $K$ decay (up to a factor 3), compared to the NP effects in the rare $B$ decay (up to $15\%$), the correlation of models with MFV is strongly broken. This plot gives a testable way to distinguish this scenario from MFV models.
 \item $K^+\rightarrow \pi^+\nu\bar\nu$ vs $S_{\psi\phi}$ (right panel): Both observables can be significantly enhanced, but not simultaneously. This particular anticorrelation can be tested by future experiments (LHCb, J-PARC,...).
 \end{itemize} 

\section{Conclusions} 
A careful analysis of the operator structure of the flavor sector of the RS model with custodial protection and a detailed fine-tuning investigation of $K$ and $B_{s,d}$ meson mixings observables show that the only stringent bound on the KK scale $M_{KK}$, coming from $\Delta F=2$ observables, is represented by $\epsilon_K$. Fixing $M_{KK}$ in the reach of LHC, once that these observables, together with SM fermion masses and mixings, are fitted, the pattern of NP deviations is clear: 

\begin{itemize}
\item The CP violating observable $S_{\psi\phi}$ can have large contributions of NP, as well as the branching ratios of the rare decays of $K$ mesons.
\item The same does not hold for the rare decays of $B$ mesons, where the NP contributions are quite small, probably below the sensitivity of LHCb.
\item A future measurement of large NP contributions in $S_{\psi\phi}$ would close the NP room for $K$ rare decays. This anticorrelation supplies a good tool to eventually put in serious difficulties the model, once more data will be available.
\end{itemize}


\vspace{-0.2cm}
\begin{theacknowledgments}
I warmly thank the other authors of~\cite{Blanke:2008zb,Buras:2009ka,Blanke:2008yr} for the very pleasant collaboration. This work was supported by the European Community's Marie Curie Research Training Network under contract MRTN-CT-2006-035505 ["HEP-TOOLS"].
\end{theacknowledgments}

\end{document}